\documentclass[12pt,aip]{article}
       \newcommand{\paraparb}[2]{\frac{\partial #1}{\partial #2}}

       \newcommand{\bxi}{\mbox{\boldmath $\xi$}}
       \newcommand{\bv}{{\mathbf v}}

       \newcommand{\bB}{{\mathbf B}}

       \newcommand{\pars}{\partial}
\begin{document}
\title{On MHD stability of gravitating plasmas with field aligned flows}
\author{H. Tasso\footnote{het@ipp.mpg.de}, G. N.
Throumoulopoulos\footnote{gthroum@cc.uoi.gr} \\
$^\star$
Max-Planck-Institut f\"{u}r Plasmaphysik \\
Euratom Association \\  85748 Garching bei M\"{u}nchen, Germany \\
$^\dag$ University of Ioannina,\\ Association Euratom-Hellenic
Republic,\\ Department of Physics, GR 451 10 Ioannina, Greece}
\maketitle

%


%

\begin{abstract}

A previous stability condition (see Throumoulopoulos and Tasso, Physics of
Plasmas 14, 122104 (2007)) for incompressible plasmas with field aligned flows
is extended to gravitating plasmas including self-gravitation. It turns out that
the stability condition is affected by gravitation through the equilibrium
values only.
\end{abstract}

\vspace{0.5cm}


\vspace{20mm}

\newpage

 The system under consideration is a constant density plasma
with an incompressible flow aligned with the magnetic field, and
maintained in a magnetohydrodynamic equilibrium by the Lorenz
force and external and self-gravitation. It is governed by the
following equations
\begin{equation}
{\bf v} = \lambda {\bf B}
\end{equation}
with $\lambda$ an arbitrary function,
\begin{equation}
{\bf \nabla}\cdot{\bf v} = 0,
\end{equation}
\begin{equation}
{\bf v}\cdot{\bf \nabla}{\bf v} = {\bf J}\times{\bf B} - {\bf \nabla}P + {\bf
\nabla}\phi,
\end{equation}
where ${\bf v}$ is the velocity, ${\bf B}$ is the magnetic field
and ${\bf J} = {\bf \nabla}\times {\bf B}$ is the current density.
$\phi$ is the gravitational potential. The equations are written
in convenient units and the constant density is set to 1. (In
fact, as will be proved below it suffices to assume  constant
 density only at equilibrium.) In addition we assume
the existence of well defined equilibrium magnetic surfaces
labelled by a smooth function $\psi$. These equations differ from
the corresponding equations of Ref.\cite{tt1} by the inclusion of
the gravitational potential.

Equations (1) and (2) lead to $\lambda = \lambda (\psi)$, and
using the identity ${\bf v}\cdot{\bf \nabla}{\bf v} = \nabla
v^{2}/2 - {\bf v}\times\nabla\times {\bf v}$, (1)-(3) lead to
\begin{equation}
(1 - \lambda^{2}){\bf J}\times{\bf B} = \nabla (P + \frac{\lambda^{2}B^{2}}{2} -
\phi) - B^{2}\nabla\frac{\lambda^{2}}{2},
\end{equation}
where $B$ and $v$ are  the magnetic field  and velocity field moduli
respectively. Taking the scalar product of Eq.(4) with ${\bf B}$ gives
\begin{equation}
P + \frac{\lambda^{2}B^{2}}{2} - \phi = P_{s\phi}(\psi),
\end{equation}
where $P_{s\phi}(\psi)$ is the arbitrary function in the presence of gravity.

Consequently, Eq.(4) can be put in the form
\begin{equation}
(1 - \lambda^{2}){\bf J}\times{\bf B} = P'_{s\phi}\nabla\psi -
(\lambda^{2})'\frac{B^{2}}{2}{\bf \nabla}\psi,
\end{equation}
with
\begin{equation}
{\bf J}\times{\bf B} = g_\phi(\psi, B^{2}){\bf \nabla}\psi,
\end{equation}
where
\begin{equation}
g_\phi(\psi, B^{2}) = \frac{P'_{s\phi}}{1 - \lambda^{2}} -
\frac{(\lambda^{2})'}{1 - \lambda^{2}}\frac{B^{2}}{2}.
\end{equation}

The only difference with Ref.\cite{tt1} is the presence of the gravitational
potential $\phi$ in (4)-(8). This potential obeys Poisson's equation
\begin{equation}
\Delta \phi = -1,
\end{equation}
whose solution is the superposition of the internal potential $\phi_{i}$ given
by $\phi_{i} = \int_V \frac{1}{|({\bf x} - {\bf x'})|}d^{3}{\bf x'}$, where $V$
is the volume of the plasma considered, and an external potential $\phi_{e}$
due to fixed masses outside the plasma.

Let us now come to the stability of our flowing and gravitating
equilibria. The general problem without gravity has been already
formulated in Ref.\cite{fr} in terms of the Lagrangian
displacement $\bxi$. Its application to equilibria with field
aligned incompressible flows and homogeneous density was
investigated in Ref.\cite{tt1} and Ref.\cite{vi}. The correct
derivation of the final condition was done in Ref.\cite{tt1}.
This condition states that a general steady state of a plasma of
uniform density and incompressible flow parallel to $\bB$ is
stable to small three-dimensional perturbations if  the flow is
sub-Alfv\'enic ($\lambda^2<1$) and $A\geq 0$, where
\begin{equation}
A = -g^{2}[(1 - \lambda^{2})({\bf J}\times\nabla\psi)\cdot ({\bf
B}\cdot\nabla)\nabla\psi + \frac{(\lambda^{2})'}{2}
|\nabla\psi|^{2}(\nabla\psi\cdot\nabla \frac{B^{2}}{2} +
g|\nabla\psi|^{2})].
\end{equation}
The quantity  $g$ is like the $g_{\phi}$ of Eq.(8) but the $P_{s}$
in it does not contain the gravitational potential $\phi$. Now,
$A\geq 0$ remains valid if $g$ in (10) is replaced by $g_{\phi}$.
The reason is that the perturbed gravity in the linearized
dynamical equations vanishes for the $\nabla\cdot{\bxi} = 0$ and
homogeneous density at equilibrium.  To prove this statement we
employ the relations\cite{fr}
 \begin{equation} \rho_1=-\nabla
 \cdot\left(\rho\bxi\right),
                             \label{1}
 \end{equation}
 \begin{equation}
 \bv_1= \paraparb{\bxi}{t}+\left(\bv\cdot\nabla
 \bxi-\bxi\cdot\nabla \bv\right),
                             \label{2}
 \end{equation}
 where the subscript 1 denotes  first order
 quantities. Acting the divergence operator on (\ref{2}) and using
 the identity
 $
 \nabla\times \left(\bxi\times
 \bv\right)=\bv\cdot\nabla\bxi-\bxi\cdot\nabla\bv  +\bxi\left(\nabla\cdot \bv\right)-
 \bv\left(\nabla\cdot \bxi\right)
 $
 with $\nabla\cdot\bv=0$ yields
 \begin{equation}
 \nabla\cdot\bv_1=\nabla\cdot\paraparb{\bxi}{t}+\bv\cdot\nabla\left(\nabla
 \cdot \bxi\right)\equiv\frac{D}{Dt}\left(\nabla
 \cdot \bxi\right).
                             \label{3}
 \end{equation}
 Therefore, if $\nabla\cdot \bxi=0$ it follows $\nabla\cdot \bv_1=0$  and,
 inversely, $\nabla\cdot \bv_1=0$ implies $\nabla\cdot \bxi=0$ for
 $t>t_0$ if this relation is satisfied initially at $t=t_0$. Then,
 (\ref{1})  in conjunction  with
 the linearized continuity equation,
 $ \pars{\rho_1}/\pars{t}+\bv\cdot\nabla\rho_1=0$, lead to
 $\rho_1=0$.
 The fact that for uniform equilibrium density,  the
 density remains uniform in the perturbed state because of
 incompressibility was not noticed in Refs.\cite{tt1} and
 \cite{vi}.

We conclude by stating that the above mentioned condition  is
sufficient  for stability if we replace
  $g$ in (10) by $g_{\phi}$ [Eq. (8)], which means that this stability
condition is affected by gravity through the equilibrium values
only.

\vspace{20mm}

\begin{center}
{\Large \bf Acknowledgements}
\end{center}

 \vspace{10mm}

Part of this work was conducted during a visit of the author G.N.T. to the
Max-Planck-Institut f\"{u}r Plasmaphysik, Garching. The hospitality of that
Institute is greatly appreciated.

This work was performed within the participation of the University of Ioannina
in the Association Euratom-Hellenic Republic, which is supported in part by the
European Union and by the general Secretariat of Research and Technology of
Greece. The views and opinions expressed herein do not necessarily reflect those
of the European Commission.

\end{document}